\documentclass[pss]{wiley2sp} 
\usepackage{amsmath}
\usepackage{amssymb}
\usepackage{widetext}
\usepackage{bm}

\newcommand{\irr}{{\mathrm{irr}}}
\graphicspath{{figures/}}

\newcommand{\psih}{\hat{\psi}}

\newcommand{\Mcal}{{\mathcal{M}}}

\newcommand{\gammah}{\hat{\gamma}}

\newcommand{\Eq}[1]{Eq.~\eqref{#1}}

\newcommand{\la}{\langle}
\newcommand{\ra}{\rangle}

\newcommand{\cop}[2]{\hat{#1}^\dagger_{#2}}

\newcommand{\aop}[2]{\hat{#1}_{#2}}

\newcommand{\bigdiagram}[1]{\includegraphics[scale=1.0]{figures/#1.pdf}}

\newcommand{\Ncal}{\mathcal{N}}
\newcommand{\Pcal}{\mathcal{P}}
\newcommand{\Qcal}{\mathcal{Q}}
\newcommand{\Ucal}{\mathcal{U}}

\newcommand{\bfx}{\mathbf{x}}

\newcommand{\Uop}{\hat{\Ucal}}

\begin{document}

\title{Diagrammatic Expansion for Positive Spectral Functions in the Steady-State Limit}

\author{%
  M. J. Hyrk\"as\textsuperscript{\Ast,\textsf{\bfseries 1}},
  D. Karlsson\textsuperscript{\textsf{\bfseries 1}},
  R. van Leeuwen\textsuperscript{\textsf{\bfseries 1}}}

\mail{e-mail
  \textsf{markku.hyrkas@jyu.fi}}

\institute{%
  \textsuperscript{1}\,Department of Physics,
Nanoscience Center P.O.Box 35 FI-40014 University of Jyv\"{a}skyl\"{a}, Finland}

\keywords{Non-equilibrium Green's functions, perturbation theory, spectral properties}

\abstract{\bf%
Recently, a method was presented \cite{Stefanucci2014} for constructing self-energies within many-body perturbation theory that is guaranteed to produce a positive spectral function for equilibrium systems, by representing the self-energy as a product of half-diagrams on the forward and backward branches of the Keldysh contour. We derive an alternative half-diagram representations that is based on products of retarded diagrams. Our approach extends the method to systems out of equilibrium. When a steady-state limit exists, we show that our approach yields a positive definite spectral function in the frequency domain.}

\maketitle

\section{Introduction}
\label{sec:1_introduction}

In Ref.~\cite{Stefanucci2014}, Stefanucci et. al. derive a diagrammatic method for generating approximations for the self-energy that are guaranteed to produce positive semidefinite (PSD) spectral functions for equilibrium systems. It was shown that such approximate self-energies can be expressed as products of half-diagrams. These results were further applied to response functions in Ref.~\cite{Uimonen2015}. The approach of Ref.~\cite{Stefanucci2014} is based on deriving a Lehmann-like representation for the correlation self-energy that in effect splits the Keldysh contour between the forward and backward branches partitioning the self-energy diagrams into time ordered and anti-time ordered half-diagrams. This approach requires the assumption that the interactions are adiabatically turned off in the future, which restricts the method to systems in equilibrium.

Below we will present an alternative formulation of the method in which the adiabatic turn-off in the future is avoided. This allows for the derivation of a Lehmann-like representation for the correlation self-energy that is valid also out of equilibrium~\cite{Gramsch2013b,Gramsch2015}. In this formulation the self-energy diagrams are partitioned into two retarded pieces.

A special non-equilibrium situation emerges when after application of an  external potential the system reaches a steady-state in the distant future. A commonly studied case is that of steady current in quantum transport, which is reached after the application of a bias. However, we may envisage many other situations, such as the attainment of a steady photocurrent of an illuminated solid, or persistent currents after application of a magnetic field in a spatially periodic system. Other examples can be conceived of when external fields couple to, e.g., the spin degrees of freedom in a system. In these cases the steady-state limit implies that we recover time-translational invariance in the long-time limit. If a steady-state is reached, our method proves that for the exact case, the spectral function is positive semidefinite (PSD) in the frequency domain. A general diagrammatic approximation will violate the PSD property~\cite{Stefanucci2014}. The method of repairing the PSD property by a minimal addition of diagrams is the same in our extension as in Ref.~\cite{Stefanucci2014}.

We begin by briefly presenting the theoretical context, and then derive the Lehmann-like representation for the correlation self-energy following the example of Ref.~\cite{Stefanucci2014} with only minor modifications. In the subsequent section we rewrite the representation in terms of explicitly retarded diagrams. In the final section, we consider the $GW$ approximation as an example, and show that it gives PSD spectral functions in the steady-state limit.


\section{Theoretical Background}

We consider an interacting fermion system described by a Hamiltonian of the form
\begin{equation} \begin{split} \label{eq:hamiltonian}
\hat{H}(t) &= \int d\bfx \cop{\psi}{}(\bfx) h(\bfx,t) \aop{\psi}{}(\bfx) \\
&+ \frac{1}{2}\int d\bfx d\bfx'\cop{\psi}{}(\bfx) \cop{\psi}{}(\bfx') v(\bfx,\bfx') \aop{\psi}{}(\bfx') \aop{\psi}{}(\bfx).
\end{split} \end{equation}
The operators  $\psi(x)$ $(\psi^\dagger(x))$ are annihilation (creation) field operators in space-spin point $\bfx$. The term $h(\bfx,t)$ is a general time-dependent one-body part, while $v(\bfx,\bfx')$ is a general two-body interaction.

The single-particle Green's function is defined as
\begin{equation}
G(\bfx_1z_1;\bfx_2z_2) = -i \la \Psi_0 | \mathcal{T}_\gamma \left\{ \psih_{H}(\bfx_1z_1) \psih^\dagger_{H}(\bfx_2z_2) \right\} | \Psi_0 \ra,
\end{equation}
where $\Psi_0$ is the initial state with $n$ particles at time $t_0$, $z_1$ and $z_2$ are time-parameters on the Keldysh contour $\gamma$:
\begin{equation}
\bigdiagram{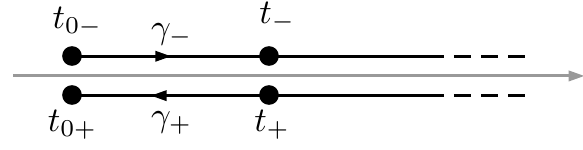}
\end{equation}
and the Heisenberg operators are given by
\begin{equation}
\aop{\psi}{H}(\bfx t) = \Uop(t_0, t) \aop{\psi}{}(\bfx) \, \Uop(t,t_0),
\end{equation}
where $\Uop(t,t_0)$ is the time-evolution operator~\cite{Stefanucci2013}.
The irreducible correlation self-energy can be expressed as \cite{Danielewicz1984,Stefanucci2013}
\begin{equation} \begin{split} \label{eq:sigmac}
&\Sigma_c(\bfx_1z_1;\bfx_2z_2) \\
&= -i \la \Psi_0 | \mathcal{T}_\gamma \left\{ \aop{\gamma}{H}(\bfx_1z_1) \cop{\gamma}{H}(\bfx_2z_2) \right\} | \Psi_0 \ra_{\irr},
\end{split} \end{equation}
with
\begin{equation} \begin{split}
\aop{\gamma}{}(\bfx_1) &= \int d\bfx_2 v(\bfx_1,\bfx_2) \aop{n}{}(\bfx_2) \aop{\psi}{}(\bfx_1),
\end{split} \end{equation}
and similarly for the adjoint $\aop{\gamma}{}^\dagger(\bfx_1)$.
The subscript $\irr$ denotes that all reducible diagrams (those that can be separated into two disjoint pieces by removing a single Green's function line) are to be removed from the expansion.

\section{Lehmann Representation of The Self-Energy}

We derive a Lehmann-like representation for the correlation part of the interaction self-energy $\Sigma_c$, following closely Ref.~\cite{Stefanucci2014}. The idea is to obtain an expression for $\Sigma_c$ that consists of a sum of squared amplitudes. From this the PSD property of the resulting spectral function can be derived in the steady-state case.

The lesser component of the correlation self-energy of \Eq{eq:sigmac} is given by
\begin{equation} \begin{split} \label{eq:sigmac_lesser}
\Sigma_c^<(1;2) = &i \left[ \la \Psi_0 | \cop{\gamma}{H}(2) \aop{\gamma}{H}(1) | \Psi_0 \ra \right]_{\irr} \\
= &i \Big[ \la \Psi_0 | \Uop(t_0,t_2) \cop{\gamma}{}(\bfx_2) \Uop(t_2,t_0) \\
&\times \Uop(t_0,t_1) \aop{\gamma}{}(\bfx_1) \Uop(t_1, t_0) | \Psi_0 \ra \Big]_{\irr}, 
\end{split} \end{equation}
where we use the shorthand notation $1 = \bfx_1t_1$, and similarly for the primed argument. The treatment of the greater component is analogous.

To proceed, we consider a complete set of states $|\chi_i \ra$ in Fock space and insert the unit operator
\begin{equation}
\mathbf{1} = \sum_i | \chi_i \ra \la \chi_i |
\end{equation}
between $\gammah_H$ and $\gammah^\dagger_H$ in \Eq{eq:sigmac_lesser}. Since $\gammah_H$ ($\gammah^\dagger_H$) removes (adds) a particle, we can restrict the sum over states to $(n-1)$--particle states. This yields
\begin{equation} \begin{split} \label{eq:split_1}
\Sigma_c^<(1;2) 
= &i \Big[ \sum_{i} S_i(2) S^*_i(1) \Big]_{\irr},
\end{split} \end{equation}
where we defined the amplitudes
\begin{equation} \label{eq:S_i}
S_i(1) = \la \Psi_0 | \Uop(t_0, t_1) \cop{\gamma}{}(\bfx_1) \Uop(t_1,t_0) | \chi_i \ra. 
\end{equation}
In \Eq{eq:split_1} the expression inside the square brackets is a Lehmann-like representation for the lesser component of reducible self-energy. To obtain a Lehmann-like representation for the irreducible self-energy $\Sigma_c$ we derive the diagrammatic representation of the amplitudes $S_i$.

To do this, we again follow the approach of Ref.~\cite{Stefanucci2014}. We assume that the initial state $|\Psi_0 \ra$ is the ground state of the system described by \Eq{eq:hamiltonian} at $t_0$. A necessary condition for having a diagrammatic expansion is the ability to use the Wick theorem. For simplicity, in this work we employ the Gell-Mann and Low theorem~\cite{Fetter2003} to connect the interacting state to a non-interacting state $| \Phi_0 \ra$ at time $-\tau$, for which the limit $\tau \to \infty$ is taken at the end. This implies that the ground state can be obtained by
\begin{equation}
|\Psi_0 \ra = \Uop(t_0,-\tau) | \Phi_0 \ra.
\end{equation}
Here $\mathcal{U}(t_0,-\tau)$ is the time-evolution operator that contains an adiabatically switched two-body interaction $v(\bfx,\bfx',t) = \mathrm{e}^{\eta (t-t_0)} v(\bfx,\bfx') $ where  $\eta$ is a positive infinitesimal taken to be zero at the end. Under the adiabatic assumption, the amplitudes $S_i$, \Eq{eq:S_i}, can be written as
\begin{equation} \label{SAmplitude1}
S_i(2) = \la \Phi_0 | \Uop(-\tau, t_2) \cop{\gamma}{}(\bfx_2) \Uop(t_2,-\tau) | \chi_i \ra. 
\end{equation}
Since $\cop{\gamma}{}$ creates one particle, only the states with one particle less than $| \Phi_0\ra$ contribute to $S_i$. Because $| \Phi_0\ra$ is a non-interacting state, a complete basis can be constructed through
\begin{equation} \label{eq:basis_generation}
| \chi_{\Pcal\Qcal}^{(N)} \ra = \cop{c}{q_N} \ldots \cop{c}{q_1} \aop{c}{p_{N+1}} \ldots \aop{c}{p_1} | \Phi_0 \ra, 
\end{equation}
where $\Pcal = \{ p_1, \ldots, p_{N+1} \}$ and $\Qcal = \{ q_1, \ldots, q_N \}$ are lists of one-particle eigenstates of the  Hamiltonian of the non-interacting system at $-\tau$. The operators $\cop{c}{k}$ and $\aop{c}{k}$ creates particles and holes in the one-particle states with quantum label $k$. $N$ is the number of particle-hole pairs created on top of the single-hole state $\hat{c}_{p_1} | \Phi_0 \ra$. The states $| \chi_{\Pcal\Qcal}^{(N)} \ra$ differ from the states $| \chi_{i} \ra$ by a different normalization~\cite{Stefanucci2014}
\begin{equation}
\la \chi_{\Pcal\Qcal}^{(N)} | \chi_{\Pcal\Qcal}^{(N)} \ra = N! (N+1)!,
\end{equation}
and we therefore need to make the replacement
\begin{equation} \label{eq:unity_operator}
\sum_i | \chi_i \ra \la \chi_i | \rightarrow \sum_{N=0}^\infty \frac{1}{(N + 1)!N!} \sum_{\Pcal \Qcal} | \chi_{\Pcal\Qcal}^{(N)} \ra \la \chi_{\Pcal\Qcal}^{(N)} |.
\end{equation}
where the sum is over all the different lists of quantum numbers that denote either an unoccupied ($\Pcal$) or an occupied ($\Qcal$) state. The prefactor is needed since different permutations of the same quantum numbers in $\Pcal$ and $\Qcal$ produce the same state (up to a minus sign that gets canceled out in \Eq{eq:unity_operator}). The replacement in \Eq{eq:unity_operator} leads to the expression
\begin{equation} \begin{split} \label{eq:self_energy_product_1}
&\Sigma_c^<(1;2) \\
&= i \Bigg[ \sum_{N=0}^\infty \frac{1}{(N + 1)!N!} \sum_{\Pcal \Qcal} S_{N,\Pcal\Qcal}(2) S^*_{N,\Pcal\Qcal}(1) \Bigg]_{\irr},
\end{split} \end{equation}
with the amplitude
\begin{equation} \begin{split}
S_{N,\Pcal\Qcal}(2) &= \la \Phi_0 | \Ucal(-\tau,t_2) \cop{\gamma}{}(\bfx_2) \Ucal(t_2,-\tau) | \chi_{\Pcal\Qcal}^{(N)} \ra.
\end{split} \end{equation}
Writing the state $| \chi_{\Pcal\Qcal}^{(N)} \ra$ as in \Eq{eq:basis_generation} leads to
\begin{widetext}
\begin{equation} \begin{split} \label{eq:contour_ordered_amplitude}
S_{N,\Pcal\Qcal}(2) &= \la \Phi_0 | \Ucal(-\tau,t_2) \cop{\gamma}{}(\bfx_2) \Ucal(t_2,-\tau) \cop{c}{q_N} \ldots \cop{c}{q_1} \aop{c}{p_{N+1}} \ldots \aop{c}{p_1} | \Phi_0 \ra \\
&= \la \Phi_0 | \mathcal{T}_{\gamma_\tau} \left\{ e^{-i \int_{\gamma_\tau} d\bar{z} \hat{H}(\bar{z})} \cop{\gamma}{}(\bfx_2 z_2) \cop{c}{q_N}(-\tau_-) \ldots \cop{c}{q_1}(-\tau_-) \aop{c}{p_{N+1}}(-\tau_-) \ldots \aop{c}{p_1}(-\tau_-) \right\} | \Phi_0 \ra,
\end{split} \end{equation}
\end{widetext}
where the contour ordering is now over the extended contour $\gamma_\tau$:
\begin{equation}
\bigdiagram{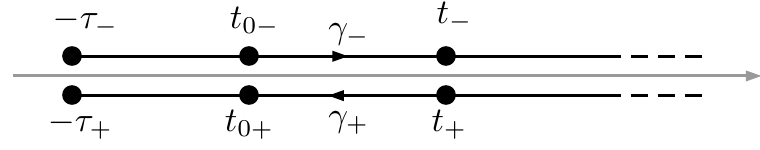}
\end{equation}
and the creation and annihilation operators have been given a time-argument to mark their position at the beginning of the contour. The contour-ordered expression, \Eq{eq:contour_ordered_amplitude}, is proportional to a $(N+2)$--particle Green's function, and can thus be diagrammatically expanded using standard perturbation theory with the Wick's theorem.

\begin{figure*}
\centering
\includegraphics[scale=1.4]{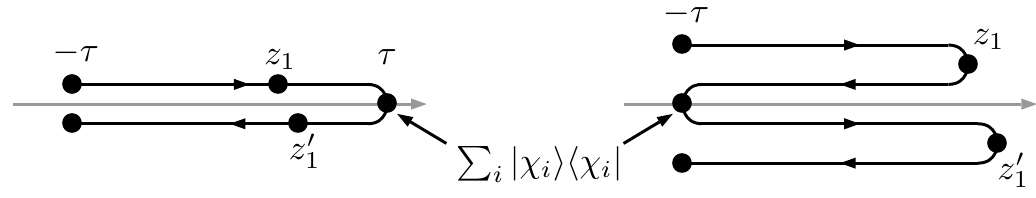}
\caption{Left figure represents the approach of \cite{Stefanucci2014}, where the unit operator can be thought to be placed at the end-point of the contour, splitting it into a time ordered forward branch and an anti-time ordered backward branch. In contrast we place the unit operator at $-\tau$, leaving a full Keldysh contour on both sides. \label{fig:comparison}}
\end{figure*}


Next we will need to define a modified $S$ in such a way that the product in \Eq{eq:self_energy_product_1} will generate only the irreducible diagrams. Here the situation is  analogous to that in \cite{Stefanucci2014}, where it was argued that this can be done by a) leaving out the term $N=0$, that contains only reducible diagrams, by starting the sum from $N=1$ and b) including only those diagrams in $S_{N,\Pcal\Qcal}(1)$ that are irreducible in the sense that the vertex $1$ can not be detached from the vertices specified by $\Pcal$ and $\Qcal$ by removing a single Green's function line. The part of $S$ that is irreducible in this sense will be denoted by $\tilde{S}$. This allows the lesser self-energy to be written as
\begin{equation} \label{eq:lehmann_representation}
\Sigma_c^<(1;2) = i \sum_{N=1}^\infty \frac{1}{(N + 1)!N!} \sum_{\Pcal \Qcal} \tilde{S}_{N,\Pcal\Qcal}(2) \tilde{S}^*_{N,\Pcal\Qcal}(1),
\end{equation}
which can be seen as a Lehmann-like representation for the non-equilibrium irreducible correlation self-energy.

It was shown in \cite{Stefanucci2014} that the Fourier transform of $-i\Sigma_c$ obtained from such a representation will be PSD in equilibrium, and that therefore the resulting spectral function will be PSD as well. The same proof can be used without modifications in the more general steady-state case. This shows that the spectral function will be PSD in the steady-state limit. 

For every diagram $D^{(j)}_{N,\Pcal\Qcal}$ in the expansion of $\tilde{S}_{N,\Pcal\Qcal}$, the expansion also contains all the diagrams that are obtained from $D^{(j)}_{N,\Pcal\Qcal}$ by permuting $\Pcal$ and $\Qcal$ (with a minus sign for odd permutations), forming a subset of related diagrams. Let $D^{(j)}_{N,\Pcal\Qcal}$ for $j \in I_N$ form a set that contains a single diagram from each such subset. We can then rebuild $\tilde{S}$ by summing over permutations of the diagrams $D^{(j)}$:
\begin{equation}
\tilde{S}_{N,\Pcal\Qcal} = \sum_{j \in I_N} \sum_{\substack{P_p \in \pi_{N+1} \\ P_q \in \pi_N}} (-1)^{|P_p| + |P_q|} D^{(j)}_{N,P_p(\Pcal)P_q(\Qcal)},
\end{equation}
where $\pi_N$ is the symmetric group of order $N$ and $|P_p|$ is the number of transpositions in the permutation $P_p$. Now in the product of $\tilde{S}$'s permuting the quantum numbers in both factors in the same way always results in the same diagram. This leads to the same diagram appearing $(N+1)!N!$ times, and consequently $\Sigma^<$ can be expressed as
\begin{equation} \begin{split} \label{eq:D_representation}
\Sigma_c^<(1;2) = &i \sum_{N=1}^\infty \sum_{j_1, j_2 \in I_N} \sum_{\substack{P_p \in \pi_{N+1} \\ P_q \in \pi_N}} (-1)^{|P_p| + |P_q|} \\
&\times \sum_{\Pcal \Qcal} D^{(j_2)}_{N,\Pcal\Qcal}(2) D^{(j_1)^*}_{N,P_p(\Pcal)P_q(\Qcal)}(1),
\end{split} \end{equation}
where the sum is only over the relative permutations between the two factors. This representation  is useful for constructing PSD approximations. 

\Eq{eq:lehmann_representation} and \Eq{eq:D_representation} are closely related to the similar equations derived in \cite{Stefanucci2014}. We will now clarify the difference between our derivations. In \cite{Stefanucci2014} it is assumed that evolving the non-interacting ground state from $-\tau$ to $\tau$ produces the same state up to a phase factor, so that
\begin{equation} \label{eq:past_and_future_states}
\Ucal(\tau, -\tau) |\Phi_0\ra = e^{i\alpha}|\Phi_0\ra.
\end{equation}
This fact is used to write the amplitude $S_i$ as
\begin{equation} \begin{split}
S_i(2) = \la \Phi_0 | \Uop(-\tau, t_2) \cop{\gamma}{}(\bfx_2) \Uop(t_2,\tau) | \chi_i \ra,
\end{split}  \end{equation}
so that the basis $\chi_i$ is constructed from a non-interacting ground state in the distant future. This allows $S_i$ to be written as a time ordered product, so that $S_i^*$ becomes correspondingly an anti-time ordered product. One can think that placing the unit operator between the $\hat{\gamma}$ operators in effect splits the contour in two. In \cite{Stefanucci2014} the unit operator is placed at the end of the contour at time $\tau$, splitting it into a time ordered forward branch and an anti-time ordered backward branch. On the other hand in this paper we deform the contour by having it return to $-\tau$ between the $\hat{\gamma}$ operators, and place the unit operator at time $-\tau$ leaving a Keldysh contour with forward and backward branches on both sides (see figure \ref{fig:comparison}). Thus we avoid having to assume \Eq{eq:past_and_future_states}, and pay the price in having to treat $S$ as an object on the full contour.

Since we do not assume \Eq{eq:past_and_future_states}, we are not restricted to equilibrium situations, and \Eq{eq:lehmann_representation} and \Eq{eq:D_representation} are valid also out of equilibrium. We stress that the discussion of PSD properties of spectral functions only applies when $\Sigma_c(t_1,t_2)$ depends only on the time difference $t_1 - t_2$. This is the case in equilibrium, as well as in the steady state limit.

\section{Evaluation of the Half-Diagrams}

\begin{figure}
\centering
\includegraphics[scale=1.0]{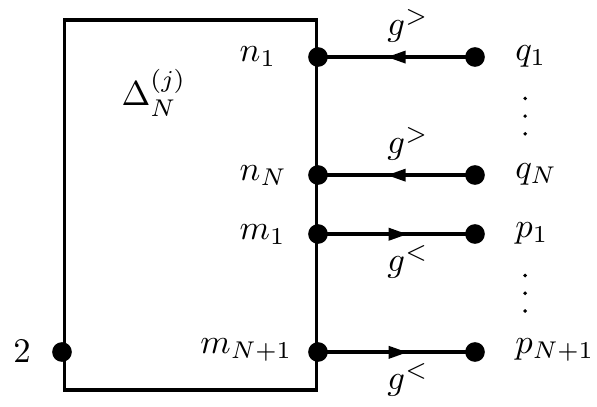}
\caption{Diagrammatic representation of the half-diagram $D^{j}_{N,\Pcal\Qcal}$ (see \Eq{eq:explicit_g_lines}). \label{fig:D_N_diagram}}
\end{figure}

Considering now a half-diagram $D^{(j)}_{N,\Pcal\Qcal}$ appearing in \Eq{eq:D_representation}, there are no interaction lines connecting to the vertices marked by $\Pcal$ and $\Qcal$, and therefore these vertices are always connected to the rest of the diagram only by a Green's function line (see figure \ref{fig:D_N_diagram}). Since the contour-times of the vertices $\Pcal$ and $\Qcal$ are always at the beginning of the contour ($-\tau_-$) these Green's functions are always lesser for $\Pcal$ and greater for $\Qcal$. Let us index the vertices that  $\Qcal$ and $\Pcal$ connect to using $\Ncal = \{ n_1,\ldots, n_N \}$ and $\Mcal = \{ m_1,\ldots, m_{N+1} \}$ respectively, and denote the Green's functions by $g^>_{n_1q_1} = g^>_{\bfx_{n_1} q_1}(t_{n_1},-\tau)$ and $g^<_{p_1m_1} = g^<_{p_1 \bfx_{m_1}}(-\tau,t_{m_1})$. We can then express $D^{(j)}_{N,\Pcal\Qcal}$ of \Eq{eq:D_representation} as
\begin{equation} \begin{split} \label{eq:explicit_g_lines}
&D^{(j)}_{N,\Pcal\Qcal}(\bfx_2z_2) \\
&= \int d\bfx_\Ncal d\bfx_\Mcal \int_\gamma dz_\Ncal dz_\Mcal \Delta^{(j)}_N(\bfx_2 z_2,\bfx_\Ncal z_\Ncal,\bfx_\Mcal z_\Mcal) \\
&\times g^>_{n_1 q_1} \cdots g^>_{n_N q_N} g^<_{p_1 m_1} \cdots g^<_{p_{N+1} m_{N+1}},
\end{split} \end{equation}
where $\Delta^{(j)}_N$ is the diagram that is left after removing the external Green's function lines from $D^{(j)}_{N,\Pcal\Qcal}$ (see figure \ref{fig:D_N_diagram}).

In contrast to \cite{Stefanucci2014} the diagrams appearing in $D$ are contour ordered rather than (anti)time ordered. To convert the contour expression into a real-time expression the usual Langreth \cite{Langreth1976} rules are inadequate due to the multi-integral structure, and more generalized rules have to be used \cite{Danielewicz1990,Hyrkas2018}. Here we give a brief discussion of the real-time conversion.

Let $A(z_\Ncal) = A(z_{n_1},\ldots,z_{n_N})$ be an arbitrary diagram containing at most two-point contour functions. We define a \emph{contour-ordered component}
\begin{equation}
A^{P(n_1) \cdots P(n_N)}(t_\Ncal),
\end{equation}
with $P$ some permutation of $\Ncal = \{ n_1, \ldots, n_N \}$, as the real-time diagram that is obtained by replacing each contour function $F(z_{n_i}, z_{n_j})$ by the greater (lesser) component if $n_i$ is left (right) of $n_j$ in the sequence $P(n_i)\cdots P(n_N)$. For example if
\begin{equation}
A(z_a, z_b, z_c) = G(z_a, z_b)G(z_b, z_c),
\end{equation}
then
\begin{equation}
A^{acb}(t_a, t_b, t_c) = G^>(t_a,t_b) G^<(t_b,t_c).
\end{equation}
Now to clean up the notation we introduce the following definitions
\begin{itemize}
\item $\Theta_{n_1 \cdots n_N} = \Theta(t_{n_1} - t_{n_2})\Theta(t_{n_2} - t_{n_3}) \cdots \Theta(t_{n_{N-1}} - t_{n_N})$ is a product of step-functions
\item A sum of contour-ordered components can be written using a sum of sequences in the superscript, as in
\begin{equation}
A^{acb} - A^{abc} = A^{acb - abc}.
\end{equation}
For brevity we use the commutator notation in this context, so that for example the above expression could be written as
\begin{equation}
A^{acb - abc} = A^{a[c,b]}.
\end{equation}
\item $[n_1,\ldots,n_N]$ denotes a nested commutator

$[\cdots[[n_1,n_2],n_3],\ldots,n_N]$.
\item A retarded component of a diagram $A(z_\Ncal)$, in which all the other arguments are retarded with respect to $t_{n_1}$, is defined as
\begin{equation} \begin{split} \label{eq:retarded_component_definition}
&A^{R(n_1,n_2 \cdots n_N)}(t_\Ncal) \\
&= \sum_{P \in S_{N-1}} \Theta_{n_1 P(n_2)\cdots P(n_N)} A^{[n_1,P(n_2),\ldots,P(n_N)]}(t_\Ncal),
\end{split} \end{equation}
where the sum is over permutations of indices other than $n_1$, so that $t_{n_1}$ is always the largest of the time-arguments. For a two-point function this definition reduces to
\begin{equation} \begin{split}
&A^{R(a,b)}(t_a,t_b) \\
&= \Theta_{a b} A^{[a,b]}(t_a,t_b) \\
&= \Theta(t_a - t_b) \left( A^>(t_a,t_b) - A^<(t_a,t_b) \right),
\end{split} \end{equation}
which coincides with the usual definition of $A^R$. Note that $A^{R(b,a)} = A^A$ is the advanced component.
\end{itemize}

An integral over all but one variables of a contour-diagram
\begin{equation}
A'(z_i) = \int_\gamma dz_{\Ncal \setminus i } A(z_\Ncal),\quad i \in \Ncal
\end{equation}
is a function symmetric with respect to the branch index, so that $A'(t) = A'(t_\pm)$, that for both branch-indices is equal to the real-time integral \cite{Danielewicz1990,Hyrkas2018}
\begin{equation} \begin{split} \label{eq:retarded_result}
A'(t_i) 
&= \int_{t_0}^\infty dt_{\Ncal \setminus i } A^{R(i,\Ncal \setminus i)}(t_\Ncal).
\end{split} \end{equation}
This result can be derived by splitting the domain of integration into sub-domains of fixed contour order. In each such sub-domain one can replace $A(z_\Ncal)$ by a specific contour ordered component. It turn out the various terms generated can be expressed elegantly using nested commutators, which motivates the definition of a general retarded component given in \Eq{eq:retarded_component_definition}. For a detailed derivation, see section 4 in \cite{Hyrkas2018}.

Applying \Eq{eq:retarded_result} to \Eq{eq:explicit_g_lines} tells us that $D^{(j)}_{N,\Pcal\Qcal}(\bfx_1z_1)$ is symmetric with respect to the branch index, and can be expressed as
\begin{equation} \begin{split}
D^{(j)}_{N,\Pcal\Qcal}(\bfx_2t_2) &= \int d\Ncal d\Mcal \Delta^{(j)R(1,\Ncal\Mcal)}_{N,\Ncal\Mcal}(2) \\
&\times g^>_{n_1 q_1} \cdots g^>_{n_N q_N} g^<_{p_1 m_1} \cdots g^<_{p_{N+1} m_{N+1}},
\end{split} \end{equation}
where $\int d\Ncal = \int d\bfx_\Ncal \int_{t_0}^\infty dt_\Ncal$ and 
\begin{equation}
\Delta^{(j)R(2,\Ncal\Mcal)}_{N,\Ncal\Mcal}(2) = \Delta^{(j)R(2,\Ncal\Mcal)}_{N}(\bfx_2 t_2,\bfx_\Ncal t_\Ncal,\bfx_\Mcal t_\Mcal)
\end{equation}
is the retarded component of the diagram $\Delta^{(j)}_N$ in which all the other arguments, including all the internal arguments, are retarded with respect to $t_2$.
The same can be done for $D^{(j)^*}$, and the result is a diagrammatic expansion for $\Sigma^<_c$ in terms of two retarded pieces that are connected by greater and lesser Green's functions. These connecting Green's functions are always either two greater or two lesser Green's functions in line in the form
\begin{equation}
\sum_q g^<_{\bfx_1 q}(t_1,-\tau) g^<_{q\bfx_2}(-\tau,t_2).
\end{equation}
These can be joined to a single Green's function by using
\begin{equation} \begin{split} \label{eq:g_split}
\sum_q g^>_{\bfx_1 q}(t_1,-\tau) g^>_{q\bfx_2}(-\tau,t_2) &= -i g^>_{\bfx_1 \bfx_2}(t_1,t_2) \\
\sum_q g^<_{\bfx_1 q}(t_1,-\tau) g^<_{q\bfx_2}(-\tau,t_2) &= i g^<_{\bfx_1 \bfx_2}(t_1,t_2).
\end{split} \end{equation}
These relations can be proven in the following way. The lesser Green's function can be written as \cite{Stefanucci2013} (the procedure for the greater component is analogous)
\begin{equation}
g^<_{\bfx q}(t_1,t_2) = i \sum_{i=1}^n \phi_{\bfx i}(t_1) \phi^*_{iq}(t_2),
\end{equation}
where $n$ is the number of particles so that the sum is over the occupied single-particle states $\phi$ (we are assuming zero temperature), non-interacting Green's functions fulfill the relation
\begin{equation} \begin{split}
&\sum_q g^<_{\bfx_1 q}(t_1,-\tau) g^<_{q\bfx_2}(-\tau,t_2) \\
&= -\sum_{i,j}^n \phi_{\bfx_1 i}(t_1) \sum_q \left[ \phi^*_{iq}(-\tau) \phi_{qj}(-\tau) \right] \phi^*_{j\bfx_2}(t_2) \\
&= -\sum_i^n \phi_{\bfx_1 i}(t_1) \phi^*_{i\bfx_2}(t_2) = i g^<_{\bfx_1 \bfx_2}(t_1,t_2).
\end{split} \end{equation}
The relations \eqref{eq:g_split} are a generalization of the equilibrium results found in \cite{Stefanucci2014}.

This joining leads ultimately to the expression
\begin{widetext}
\begin{equation} \begin{split} \label{eq:R_representation}
\Sigma_c^<(1;2) &= i \sum_{N=1}^\infty \sum_{j_1,j_2 \in I_N} \sum_{\substack{P_n \in \pi_N \\ P_m \in \pi_{N+1}}} (-1)^{|P_n| + |P_m|} \int d\Ncal d\Ncal' d\Mcal d\Mcal' \\
&\times \Delta^{(j_2)R(2,\mathcal{N}'\mathcal{M}')}_{N,\Ncal'\Mcal'}(2) g^>_{n'_1 P_n(n_1)} \cdots g^>_{n'_N P_n(n_N)} g^<_{P_m(m_1) m'_1} \cdots g^<_{P_m(m_{N+1}) m'_{N+1}} \left[\Delta^{(j_1)R(1,\Ncal\Mcal)}_{N,P_n(\Ncal)P_m(\Mcal)} \right]^*(1),
\end{split} \end{equation}
\end{widetext}
where $g^\lessgtr_{n'_1 n_1} = g^\lessgtr(\bfx_{n'_1}t_{n'_1},\bfx_{n_1}t_{n_1})$, so that the expression no longer depends on $-\tau$. \Eq{eq:R_representation} is an exact representation of the correlation self-energy in terms of retarded pieces. Furthermore, it can be used as a starting point for the repairing procedure to produce PSD self-energies that was presented in \cite{Stefanucci2014}.

A given approximate self-energy can always be written in the form of \Eq{eq:R_representation} with some $\Delta$, and some restrictions on the sums. By cutting the greater and lesser Green's function lines one can obtain an expression in the form of \Eq{eq:D_representation}, now with restricted sums. Typically such an approximation is not PSD, but it can be made PSD by addition of extra diagrams. It was shown in \cite{Stefanucci2014} that if \Eq{eq:D_representation} is modified to
\begin{equation} \begin{split} \label{eq:D_representation_approx}
\tilde{\Sigma}_c^<(1;2) = &i \sum_{N=1}^{N_{max}} \sum_{j_1, j_2 \in \tilde{I}_N} \sum_{\substack{P_p \in \tilde{\pi}_{N+1} \\ P_q \in \tilde{\pi}_N}} (-1)^{|P_p| + |P_q|} \\
&\times \sum_{\Pcal \Qcal} D^{(j_2)}_{N,\Pcal\Qcal}(2) D^{(j_1)^*}_{N,P_p(\Pcal)P_q(\Qcal)}(1),
\end{split} \end{equation}
with $\tilde{I}_N \subset I_N$, $\tilde{\pi}_N \subset \pi_N$ and $\tilde{\pi}_{N+1} \subset \pi_{N+1}$, the resulting approximate self-energy will be PSD as long as $\tilde{\pi}_N$ and $\tilde{\pi}_{N+1}$ are subgroups of the permutation groups $\pi_N$ and $\pi_{N+1}$ respectively. These observations were used in \cite{Stefanucci2014} to set out a repairing procedure for converting a non-PSD approximation to a PSD one using a minimal number of extra diagrams. These arguments apply directly also to the non-equilibrium case here discussed.

This procedure can be extended to dressed Green's functions. The discussion regarding this in \cite{Stefanucci2014} is again directly applicable to our case.

\section{The GW Approximation in the Steady-State Limit}

As an example of the results derived above, we will in this section outline the proof that the spectral functions produced by the $gW_0$ approximation in the steady state limit are PSD.

By $gW_0$ we mean the approximation in which the exchange-correlation self-energy is given by
\begin{equation} \label{eq:xc_sigma}
\Sigma_{xc,gW_0}(\bar{1},\bar{2}) = i g(\bar{1},\bar{2})W_0(\bar{1},\bar{2}),
\end{equation}
where $\bar{1} = \bfx_1z_1$,
\begin{equation}
W_0(\bar{1},\bar{2}) = V(\bar{1},\bar{2}) + \int d\bar{3}d\bar{4} V(\bar{1},\bar{3}) P(\bar{3},\bar{4}) W_0(\bar{4},\bar{2})
\end{equation}
with $V(\bar{1},\bar{2}) = \delta(z_1,z_2)v(\bfx_1,\bfx_2,z_1)$, $\int d\bar{3} = \int \bfx_3 \int_\gamma z_3$ and
\begin{equation}
P(\bar{1},\bar{2}) = -i g(\bar{1},\bar{2}) g(\bar{2},\bar{1}),
\end{equation}
the polarization function in the random phase approximation.
As explained above, to show that $gW_0$ is PSD we must show that $\Sigma^<_{c,gW_0}$ can be written in the form of \Eq{eq:D_representation_approx}.

The lesser component of \Eq{eq:xc_sigma} is (the exchange term vanishes, since it is time-local \cite{Stefanucci2013})
\begin{equation}
\Sigma^<_{c,gW_0}(1,2) = i g^<(1,2)W^<_0(1,2)
\end{equation}
and
\begin{equation}
P^<(1;2) = -i g^<(1;2)g^>(2;1).
\end{equation}
Using the equation
\begin{equation}
W_0^<(1;2) = \int d3d4 W_0^R(1;3)P^<(3;4)W_0^A(4;2)
\end{equation} 
we obtain
\begin{equation} \begin{split} \label{eq:before_cut}
&\Sigma^<_{c,gW_0}(1;2) \\
&= \int d3d4\, W_0^R(1;3) g^>(4,3) g^<(1;2) g^<(3,4) W_0^A(4;2) \\
\end{split} \end{equation}

Using the equations \eqref{eq:g_split} to cut the Green's function lines, we obtain (dropping the time-arguments for brevity)
\begin{equation} \begin{split}
&\Sigma^<_{c,gW_0}(1;2) = -i \sum_{q_1p_1p_2} \int d3d4\\
&\times W_0^R(1;3) g^>_{q_1\bfx_3} g^<_{\bfx_1p_1} g^<_{\bfx_3 p_2} g^>_{\bfx_4q_1} g^<_{p_1 \bfx_2} g^<_{p_2 \bfx_4} W_0^A(4;2) \\
\end{split} \end{equation}
Now if we expand the screened interaction as (repeated convolutions implied)
\begin{equation}
W_0^R = \sum_{j = 0}^\infty W_0^{(j)R} = \sum_{j = 0}^\infty (VP^R)^j V,
\end{equation}
with $j$ the number of polarization bubbles, we can express $\Sigma^<_c$ as
\begin{equation} \begin{split} \label{eq:gw_D_representation}
&\Sigma^<_{c,gW_0}(1;2) = -i \sum_{j_1,j_2 \in \tilde{I}_N} \sum_{\Pcal\Qcal} D^{(j_2)}_{1,\Pcal\Qcal}(2) D^{(j_1)^*}_{1,\Pcal\Qcal}(1),
\end{split} \end{equation}
with $\Pcal = \{ p_1, p_2 \}$, $\Qcal = \{ q_1 \}$, $\tilde{I}_N = \{ 0,1,2,\ldots \}$ and
\begin{equation} \begin{split}
D^{(j)}_{1,\Pcal\Qcal}(2) = \int d3\, W_0^{(j)R}(2;3) g^>_{q_1\bfx_3} g^<_{\bfx_2p_1} g^<_{\bfx_3 p_2}.
\end{split} \end{equation}
Now \Eq{eq:gw_D_representation} matches the form of \Eq{eq:D_representation_approx} for $N_{max} = 1$ and the sum over permutations including only the identity permutation. Since $\tilde{I}_N$ represents a set of diagrams not related by permutations, and since the identity permutation constitutes a sub-group by itself, it follows that $-i\Sigma^<_{c,gW_0}$ is PSD.

As mentioned, the PSD property is retained in the dressed case, meaning that the fully self-consistent $GW$ approximation is PSD as well. Indeed numerical results yield PSD spectral functions \cite{Thygesen2008,Myohanen2009,PuigvonFriesen2010}.




\section{Conclusions}

We have presented a method for obtaining approximations for the correlation self-energy that are guaranteed to result in PSD spectral functions in non-equilibrium systems in the steady-state limit. A further advantage of our approach is that, unlike the approach of \cite{Stefanucci2014}, it is not limited to correlators of two operators, but is in principle generalizable for higher-order correlators by placing a set of basis states at distant past between each operator. As an application we showed that the steady-state spectral function within the $gW_0$ approximation is PSD. A more detailed exposition will be deferred to a future publication.

\begin{acknowledgement}
D.K. acknowledges the Academy of Finland for funding under Project
No. 308697. M.H. thanks the Finnish Cultural Foundation for support R.v.L. acknowledges the Academy of Finland for funding under Project No. 317139.
\end{acknowledgement}

\bibliographystyle{pss}
\bibliography{aForPapers-ProceedingsMarkku}

\end{document}